%% file: doc.tex
\documentclass[11pt]{article}

\usepackage{fullpage}
\usepackage[affil-it]{authblk}
\usepackage[cmex10]{amsmath}
\usepackage[margin=2cm,nohead]{geometry}
\usepackage{dsfont}
\usepackage[english]{babel}
\usepackage[applemac]{inputenc}
\usepackage[T1]{fontenc}
\usepackage[active]{srcltx}
\usepackage{amsfonts}
\usepackage{amssymb}
\usepackage{color}
\usepackage{bm}
\usepackage{bbm}
\usepackage{authblk}
\usepackage{comment}
\usepackage{enumitem}
\usepackage[titletoc,title]{appendix}
\setlist[enumerate,1]{label={(\Alph*)}}
\usepackage{cite}
\usepackage{ifpdf}              
\usepackage{array}              
\usepackage{xcolor}
\usepackage{authblk}
\usepackage{lipsum}
\usepackage{float}

\ifpdf
 \usepackage[pdftex]{graphicx}
 \usepackage{epstopdf}
 \DeclareGraphicsExtensions{.eps,.jpg,.png,.pdf}
\else
 \usepackage[dvips]{graphicx}
 \DeclareGraphicsExtensions{.eps}
\fi

\newcommand{\figpath}{./figs}

\newcommand{\energiabar}[1]{\bar{\mathcal{E}}_\text{#1}}
\newcommand{\energia}[1]{\mathcal{E}_\text{#1}}
\newcommand{\ene}{\mathcal{E}}
\renewcommand{\P}[1]{P_\text{#1}}
\newcommand{\Pbar}[1]{\bar{P}_\text{#1}}
\newcommand{\T}[1]{T_\text{#1}}

\newcommand{\meang}{\bar{\gamma}}
\newcommand{\meanP}[1]{\bar{P}_\text{#1}}
\newcommand{\nc}{n_\text{c}}
\newcommand{\Nt}{N_\text{t}}
\newcommand{\Nr}{N_\text{r}}

\newcommand{\E}[1]{\mathbb{E} \left\{#1 \right\}}

\newcommand{\fig}[4]
{
  \begin{figure}[h]	
    \centering
    \resizebox{#3}{!}{\includegraphics{\figpath/#1}}
    \caption{#4 \normalsize}
    \label{#2}
  \end{figure}
}

\begin{document}
\title{A multi-layered energy consumption model for \\smart wireless acoustic sensor networks}
\author[a,b]{Gert Dekkers\thanks{For questions related to the model: gert.dekkers@kuleuven.be}}
\author[c,d]{Fernando Rosas}
\author[b]{Steven Lauwereins}
\author[b]{Sreeraj Rajendran}
\author[b]{Sofie Pollin}
\author[b]{Bart Vanrumste}
\author[b]{Toon van Waterschoot}
\author[b]{Marian Verhelst}
\author[a]{Peter Karsmakers}
\affil[a]{Department of Computer Science, KU Leuven, Belgium}
\affil[b]{Department of Electrical Engineering, KU Leuven, Belgium}
\affil[c]{Department of Mathematics, Imperial College London, UK}
\affil[d]{Department of Electrical and Electronic Engineering, Imperial College London, UK}
\maketitle

\begin{abstract}
\input{\textpath/abstract}
\end{abstract}


\input{\textpath/introduction}
\input{\textpath/overviewconsumption}
\input{\textpath/sensingconsumption}
\input{\textpath/processingconsumption}

\input{\textpath/wirelessconsumption}

\input{\textpath/modelanalysis}
\input{\textpath/appendix}

\clearpage
\bibliographystyle{IEEEtran}
\bibliography{refs}

\end{document}

%% file: text/abstract.tex

Smart sensing is expected to become a pervasive technology in smart cities and environments of the near future. These services are improving their capabilities due to 
integrated devices shrinking in size while maintaining their computational power, which can run 
diverse Machine Learning algorithms and achieve high performance in various data-processing tasks. 
One attractive sensor modality to be used for smart sensing are acoustic sensors, 
which can convey highly informative data while keeping a moderate energy consumption. Unfortunately, the energy budget of current wireless sensor networks is usually not enough to support the requirements of standard microphones. Therefore, energy efficiency needs to be increased at all layers --- sensing, signal processing and communication --- in order to bring wireless smart acoustic sensors into the market. 

To help to attain this goal, this paper introduces \textit{WASN-EM}: an energy consumption model for wireless acoustic sensors networks (WASN), whose aim is to aid in the development of novel techniques to increase the energy-efficient of smart wireless acoustic sensors. This model provides a first step of exploration prior to custom design of a smart wireless acoustic sensor, and also can be used to compare the energy consumption of different protocols. 



%% file: text/introduction.tex

\section{Introduction}

Recent advances in hardware miniaturization is enabling integrated devices containing wireless radios, processing and sensing capabilities to shrink their sizes, while their computational power is maintained or even increased \cite{Borkar2011}. This, along with a recent surge of powerful Machine Learning algorithms that can accomplish various data-driven tasks, has caused a rising interest in smart cities and environments. Such a smart environment typically uses a network of wireless sensors for acquiring information to offer smart functionality \cite{Rawat2014}. Scenarios where this have been taking place include security, smart city, health monitoring and entertainment \cite{mainwaring2002wireless,Butun2014, Erden2016, Rawat2014, Hashem2016}.

An attractive type of sensor to use in these applications are acoustic sensors (i.e. microphones). Compared to other sensors, acoustic sensors can convey highly informative data, including sounds with semantic contect (e.g. speech), noises that represent a warning (e.g. screams), sounds with intrinsic meaning in a particular environment (e.g. a water faucet running within a kitchen), etc. Detecting these informative acoustic events can be benefitial for numerous tasks, including speech recognition, surveillance and monitoring, and many others~\cite{Vacher2011A}.

In order to allow an wireless acoustic sensor network (WASN) to be easily installed, wireless battery-powered architectures are preferred to avoid extensive use of wiring \cite{Bertrand2011}. Unfortunately, this brings additional challenges as the lifetime of the devices can be compromised by the energy consumption of acoustic sensors and wireless transmission, which usually go beyond what common current sensor network architectures can provide \cite{Lauwereins2018}.

Increasing the energy efficiency of sensors can be tackled from the different layers of the processing chain, including the sensing, signal processing and communication modules~\cite{karl2007protocols}. In effect, the total amount of consumed energy depends strongly on particular parameters related to hardware dependency. Ideally, one would optimize these parameters for each particular hardware design, but in practice such approach would require tedious measurement campaings. From a signal processing point of view, energy is often reduced by means of limiting the amount of arithmetic operations. Such an approach might not always be valid due to costly memory accesses. Additionally, if the goal is to design a smart wireless acoustic sensor, it is important to know the energy contribution of each layer to motivate an optimization. 

The literature provides a number of modeling efforts about various aspects of wireless sensors networks (see \cite{anastasi2009energy} and references therein). Substantial efforts has been done in increasing the energy efficiency of the wireless communication module, ranging from 
the physical layer~\cite{shih2001physical}, multihop and routing \cite{ganesan2001highly,rosas2015energy} and network layer protocols~\cite{ye2002energy,van2003adaptive}. Regarding the processing of audio information to retrieve relevant information, deep learning has recently become popular \cite{DCASE2013,DCASE2016,DCASE2017}. Yang et al have introduced an energy estimation model to estimate the energy consumption of Deep Neural Networks (DNN) \cite{Yang2017}. The model is based on power numbers from their Eyeriss DNN accelerator chip and provides an estimate on the energy consumption given an optimized dataflow \cite{Chen2016}. The disadvantage of the model is that it is not flexible as it is limited to the bounds of that particular chip. To the best of our knowledge, no open-source model is available that cover all layers of a smart acoustic sensor.

In order to aid the design of novel configurations for allowing the increase of energy efficiency of a range of sensing devices, in this report we introduce the \textit{WASN-EM}: an Energy Model for Wireless Acoustic Sensor Networks. The goal of the model is threefold: 
\begin{itemize}[noitemsep]
\item[(a)] to bridge the gap between the machine learning and hardware community regarding (energy-efficient) design of smart wireless acoustic sensors,
\item[(b)] to be flexible to adjust to various hardware configurations, and
\item[(c)] to provide a simple and open-source software such that the community can contribute.
\end{itemize}
The model can act as a first step prior to custom design of a smart wireless acoustic sensor and provide a common ground for researchers to compare energy consumption, computational complexity and memory storage. 

In the sequel, an overview of the proposed model is provided in section \ref{sec:enconduty} as it is composed out of three separate models: sensing, processing and communication. Section \ref{sec:enconsens} elaborates on the modelling of the sensing layer which includes the microphone, power amplifier and analog-to-digital converter. Section \ref{sec:enconproc} covers the processing layer where an hardware architecture model is introduced that provides an energy consumption estimate of the arithmic operations and the accompanying memory accesses. Additionally, some common algorithms for processing audio information are introduced of which an energy consumption estimate can be obtained using the proposed hardware architecture model. Section \ref{sec:comm} elaborates on the model of the communication layer. The model covers the power amplifier and other electronic components based on a hypothetical hardware architecture along with the effects of re-transmission. The final section provides some guidelines on interesting parameters to experiment with.


%% file: text/overviewconsumption.tex

\newpage
\section{System model}
\label{sec:enconduty}

\begin{figure}[H]
	\centering
	\includegraphics[width=0.8\textwidth]{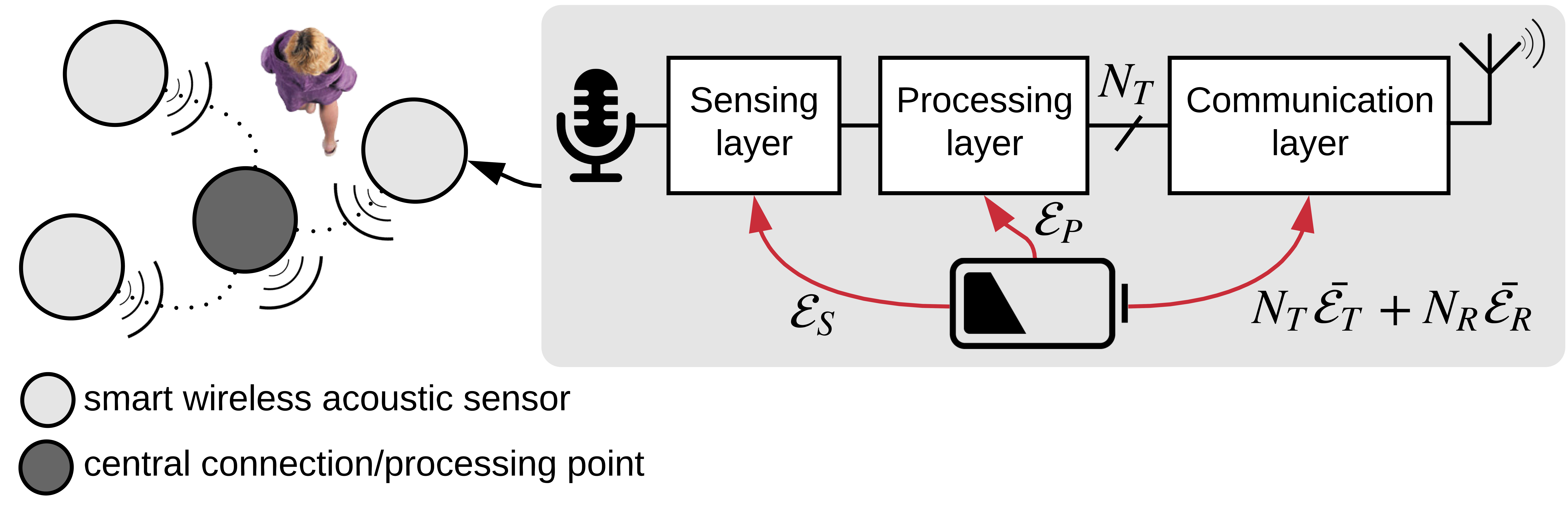}
	\caption{Scenario description}
	\label{fig:scendesc}
\end{figure}

Let us consider an scenario where the goal is to monitor a particular environment to acquire information about the activities that are taking place. This could correspond to an apartment, where by determining the activities that are taking place an automated system could optimize a range of services including lighting, heating, etc. One way to harvest information about an environment is to deploy a WASN consisting of multiple acoustic sensors nodes with wireless communication capabilities, and a central connection/processing device that can gather and process the sensed data (see Figure \ref{fig:scendesc}). Each node consists out of a acoustic sensing, processing and communication module, being capable of:
\begin{itemize}[noitemsep]
\item[(1)] capturing and digitizing acoustic information,
\item[(2)] processing the resulting acoustic data to provide a meaningfull output and/or to reduce the amount of bits to communicate, 
\item[(3)] transmitting the processed information with a central connection/processing point, and
\item[(4)] receiving data from a central connection/processing point.
\end{itemize}

Given the aforementioned scenario, let us consider a single duty cycle where a single node measures the environment during $\Delta$ seconds and subsequently does some processing on the data. Consequently, this generates $N_T$ bits of information, spending $\energia{S}$ and $\energia{P}$ joules in the sensing and data processing step respectively. The processed information is divided in $N_T/(r_\text{u}L_\text{u})$ forward frames, where $L_\text{u}$ is the number of payload bits per frame in the uplink direction and $r_\text{u}$ is its code rate, which are transmitted directly to a central connection point (sink) using designated time-slots. After each frame transmission trial, the sink send back a feedback frame which acknowledges correct reception or requests a re-transmission. Similarly, the communication module can receive $N_R/(r_\text{d}L_\text{d})$ frames where $N_R$ is total amount of received informative bits and $L_u$ and $r_u$ are the number of payload bits and the code rate in the downlink direction.  Hence, the total energy consumption of the audio sensor node can be modeled as follows:
\begin{equation}\label{eq:Etotal}
\energiabar{node} =  \energia{S} + \energia{P} +  N_T\energiabar{T} + N_R\energiabar{R}
\enspace.
\end{equation}
Above, $\energiabar{T}$ and $\energiabar{R}$ is the average total energy consumption per information bit that is correctly transmitted and respectively received.

Let us assume that the node has to be sensing the environment $\delta$ percent of the time (i.e. its duty cycle). Let us also assume that the node carries $n_\text{b}$ batteries with a charge of $\mathcal{B}$ Joules each. Then, by neglecting the energy consumption of the node when it is in sleep mode, the lifetime of the node can be estimated to be equal to
\begin{equation}\label{eq:lifetime}
\mathcal{L} =  \delta^{-1} \frac{ n_\text{b} \mathcal{B}}{ \bar{\mathcal{E}}_\text{node}} \Delta
\enspace.
\end{equation}
In the following, we will develop expressions for $\energia{S}$, $\energia{P}$ and $\energia{T}$.


%% file: text/sensingconsumption.tex

\section{Energy consumption of sensing}
\label{sec:enconsens}

The energy consumption expended in acoustic sensing can be expressed as follows:
\begin{equation}
\energia{S} = \energia{mic} + \energia{LNA,mic} + \energia{ADC,mic}
\enspace.
\end{equation}
Above, the energy consumed by the analog front-end $\energia{S}$ consists out of the energy consumed by the microphone $\energia{mic}$, the consumption of the low-noise amplifier (LNA) $\energia{LNA,mic}$  and the consumption of the analog-to-digital convertor (ADC) in digitalising the signal $\energia{ADC}$.
\fig{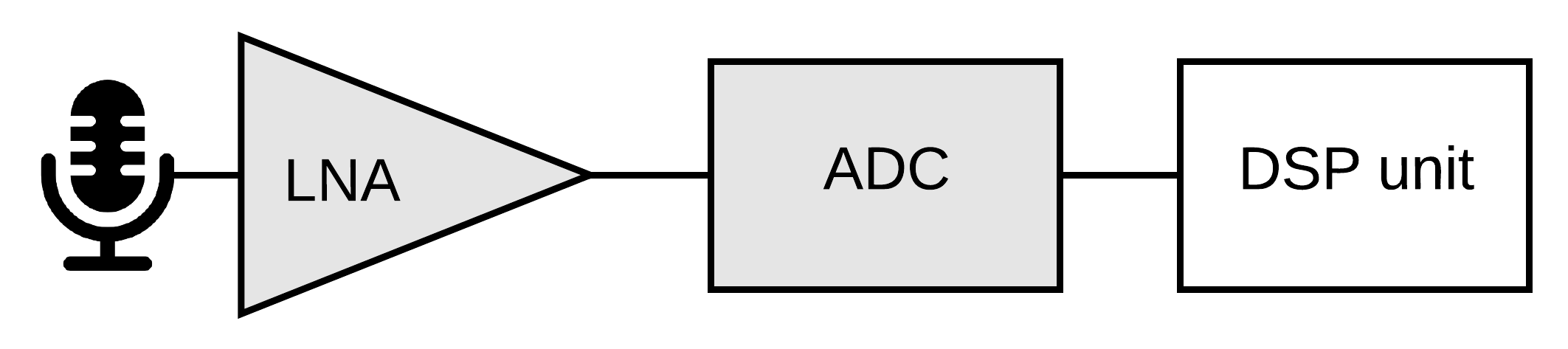}{mic}{8cm}{Microphone analog front-end (marked in grey) along with the processing layer}

The power consumption of the microphone can be expressed as follows:
\begin{equation}
\P{mic} = 
\begin{cases}
0 &\text{if passive mic or switched off},\\
\P{mic,act} &\text{if active and powered on},
\end{cases}
\end{equation}
so, the energy consumption will be given by
\begin{equation}
\energia{mic} = 
\begin{cases}
0 &\text{if passive mic, or}\\
\Delta \P{mic,act} &\text{if active.}
\end{cases}
\end{equation}

The energy consumption of the LNA can be calculated as
\begin{equation}
\energia{LNA,mic} = I_\text{LNA,mic} V_\text{dd}^\text{LNA,mic} \Delta 
\enspace.
\end{equation}
Above, $V_\text{dd}^\text{LNA,mic}$ is the voltage supply level and $I_\text{LNA,mic}$ is the average current drawn by the LNA, which can be calculated as
\begin{equation}
I_\text{LNA,mic} = \frac{\pi u_\text{T} 4kT W_\text{ADC}}{2} \left( \frac{\text{NEF}}{v_\text{n,in}^\text{rms}} \right)^2
\enspace,
\end{equation}
where $k$ is the Boltzmann constant, $T$ is the temperature in Kelvin, $u_\text{T} = kT/q_\text{e}$  with $q_\text{e}$ equal to the charge of the electron, $W_\text{ADC}$ is the ADC bandwidth, $v_\text{n,in}^\text{rms}$ is the RMS voltage of the noise at the input of the LNA, and $\text{NEF}$ is the \textit{noise efficiency factor}, which was proposed in \cite{Steyaert1987} and which value in average designs is between 5-10. Typical values for $v_\text{n,in}^\text{rms}$ are
\begin{equation}
v_\text{n,in}^\text{rms} = \begin{cases}
10 \mu W &\text{for passive microphones} \enspace,\\
100 \mu W &\text{for active microphones} \enspace.
\end{cases}\end{equation}

The energy consumption of the ADC can be computed as
\begin{equation}
\energia{ADC,mic} = \P{ADC,mic} \Delta
\enspace.
\end{equation}
Above, $\P{ADC,mic}$ is the power consumption of the ADC, which can be calculated as
\begin{equation} \label{eq:ADCconsumption}
\P{ADC,mic} = 2^{n_\text{mic}} f_\text{s,mic} \text{FOM}
\enspace,
\end{equation}
where $n_\text{mic}$ is the resolution of the ADC, $f_\text{s,mic}$ is the sampling frequency and $\text{FOM}$ is the \textit{figure of merit} of the ADC.

An overview of other relevant hardware related parameters with the used values is given in Appendix \ref{appendix:params} Table \ref{tab:paramsens}. 


%
%


%% file: text/processingconsumption.tex

\section{Energy consumption of processing}
\label{sec:enconproc}

The goal of the (optional) local processing is to translate the raw audio information to a lower dimension to reduce the amount of communicated bits. The processed information could already be the final required output (e.g. a classification output) or the output of a feature extraction stage. From a signal processing point of view, energy consumption is often reduced by means of limiting the amount of arithmetic operations. Such an approach might not always be valid due to costly memory accesses. Here the energy consumption $\energia{P}$ due processing of the acquired information is defined as

\begin{equation}\label{eq:mPconsumption}
\energia{P} = \overbrace{\energia{cc} \sum_{j=1}^{J_\text{ALU}}  c_j n_j^\text{DSP}}^{\energia{op}} + \overbrace{\Delta\sum_{k=1}^{J_\text{MEM}}(\energia{ma,k}M_{\text{a},k} + \energia{ms,k}M_{\text{s},k}}^{\energia{m}})
\enspace,
\end{equation}
which consists of the consumption due to arithmetic operations $\energia{ops}$ and due to memory $\energia{m}$. 

Regarding the energy consumed due to the arithmetic operations $\energia{op}$, $\energia{cc}$ is the energy consumption per clock cycle, $c_j$ is the number of clock cycle required by the $j$-th arithmetic operation which is performed $n_j^\text{DSP}$ times during the digital signal processing and $J_\text{ALU}$ is the number of different arithmetic operations the microprocessor performs. In the model we distinguish the following arithmetic operations: 1) multiply-accumulate (MAC), 2) addition and subtraction, 3) multiplication, 4) division, 5) comparison (including maximum and minimum), 6) natural exponentiation and 7) logarithm. Depending on the hypothetical hardware architecture (e.g. CPU, ASIC, ...), each of these operations take a different amount of clock cycles and energy cost per clock cycle. This model focusses on a microcontroller-based wireless acoustic sensor without any hardware acceleration. A model for a custom chip is not provided but in general these could provide energy gains of a factor 500 till 1000 for the processing layer \cite{Horowitz14}.

\begin{figure}[h!]
	\centering
	\includegraphics[width=0.4\textwidth]{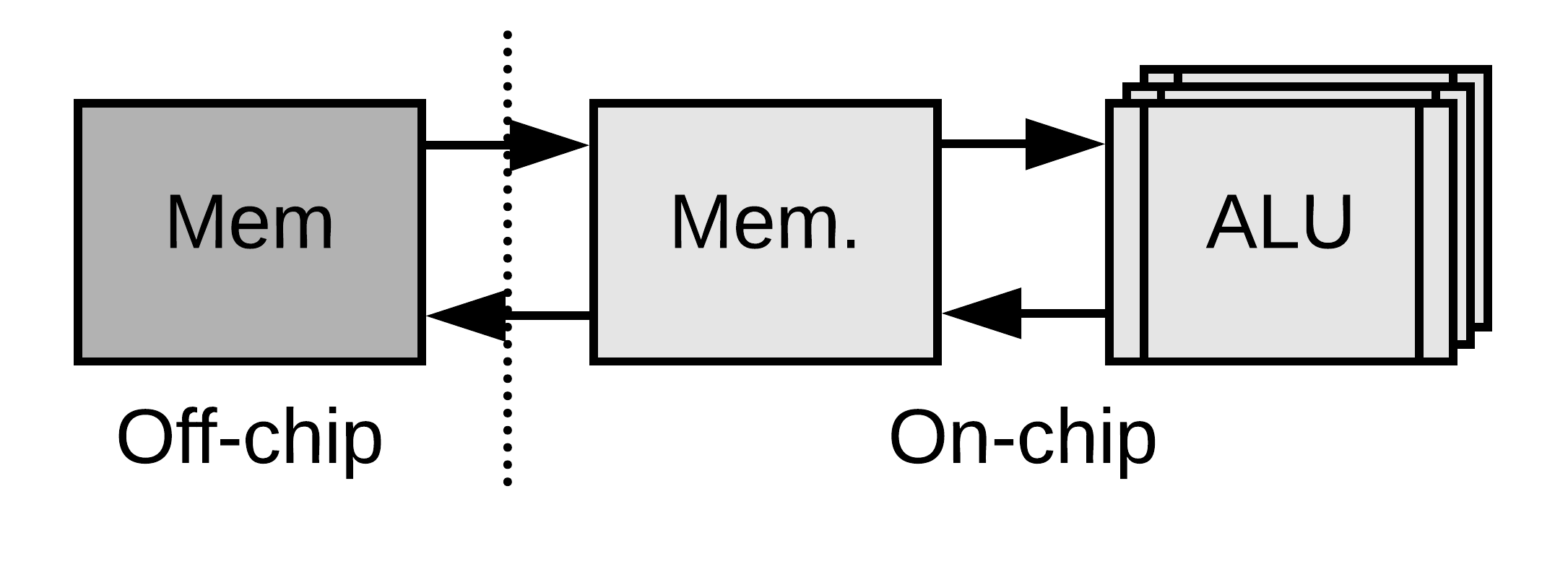}
	\caption{The hardware architecture model}
	\label{fig:hwarch}
\end{figure}

The energy consumed by the memory $\energia{m}$, is decomposed into the energy required for accessing and storing which depends on the amount of bits accessed $M_{\text{a},k}$ or stored $M_{\text{s},k}$ in a particular memory k. The energy consumed by accessing and storing one bit is defined by $\energia{ma,k}$ and storage $\energia{ms,k}$ respectively for $k=[0,1,\dots, J_\text{MEM}]$ with $J_\text{MEM}$ the amount of available memories. Typical hardware architectures have multiple memory types available each having a different energy consumption for storage and access. The least consuming memory is typically close to the processor unit but limited in size. When the needed memory is not sufficient, data movement is needed to and from more consuming memories. It is therefore important to maximize data reuse on the least consuming memories to limit data movement \cite{Sze2017}. In this model we assume: a) an architecture with on- and off-chip memory as shown by Figure \ref{fig:hwarch}, b) equal energy cost for each operation per clock cycle b) equal cost for memory read and writes and c) in-place computation such that memory accesses can easily be derived from a particular arithmetic operation. Additionally, it is explicitly defined where the information should be stored/accessed for each building block in the processing chain. An overview of the parameters of the hardware architecture model are given in Appendix \ref{appendix:params} Table \ref{tab:paramproc}.

In the following subsections the explanation and an energy model for some typical algorithms used in the field of automatic sound recognition are provided. The depicted problem is that of classifying an input audio stream in one out of all classes. A typical system to solve such a problem consists of Mel-Frequency (Cepstral) Coefficients as feature extraction followed by a (Deep) Neural Network based architecture as classifier \cite{DCASE2013,DCASE2016,DCASE2017}. Both consist of several modular building-blocks which will be described in the two following subsections. 

\subsection{Feature extraction: Mel-Frequency Cepstral Coefficients}
\begin{figure}
	\centering
	\includegraphics[width=1\textwidth]{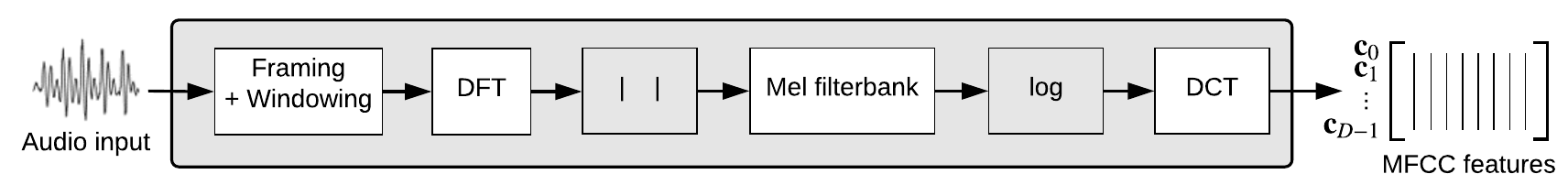}
	\caption{Mel-Frequency Cepstral Coefficients feature extraction process. The raw acoustic data is transformed to the feature domain by applying a (1) framing and windowing, (2) Discrete Fourier Transform, (3) Mel-filterbank, (4) logarithm and (5) Discrete Cosine Transform.}
	\label{fig:specAcFeat}
\end{figure}

The Mel-Frequency Cepstral Coefficients (MFCC) feature extraction algorithm originates from the domain of automatic speech recognition and is based on the perception of sound by the human auditory system \cite{Davis1980}. Despite the fact that MFCC was developed for that task, it is shown that it is also usable for automatic sound recognition due to their ability to represent the amplitude spectrum in a compact form \cite{DCASE2013,DCASE2016,DCASE2017}. Figure \ref{fig:specAcFeat} is an overview of the MFCC feature extraction process and involves the following main components: (1) framing and windowing, (2) Discrete Fourier Transform (DFT), (3) Mel-frequency filterbank, (4) logarithmic operation and (5) Discrete Cosine Transform (DCT). In recent years, related to the popularity of deep learning, researchers tend to use an intermediate output of the MFCC algorithm (Mel-frequency filterbank) or even the raw audio waveform due to the neural network being able to learn a feature representation from the provided data. Up to date, the building blocks of MFCC are still one of the dominant signal processing algorithms used for speech and audio classification tasks. In the following subsections these building blocks are explained.

\subsubsection{Framing and windowing}
The framing and windowing operation of the feature extraction process transfers the raw acoustic waveform into short overlapping segments. These segments, further called frames, are typically 30 ms long with an overlap of 10 ms. Each frame $\textbf{f}$ is then typically windowed with an Hamming-window $\textbf{h}$ to reduce spectral leakage. The windowing operation is defined as $\textbf{s}_n=\textbf{h}_n\textbf{f}_n$ for $n=[0,1,\dots,N_t-1]$ with $N_t$ the amount of samples in one frame. The amount of operations for this stage consists of $N_t$ multiply-accumulates for one frame. As it can be computed in-place, the total needed storage is $2\cdot N_t\cdot S$ bits, where $S$ represents the word size. As a multiply-accumulate needs four memory accesses, this results in a total of $4\cdot N_t$ memory accesses.

\subsubsection{Discrete Fourier transform}
The framing and windowing operation are followed by the DFT that transforms the frame $\textbf{s}$ in the time-domain into a frame $\textbf{z}$ in the frequency domain. Typically, the Fast Fourier Transform (FFT) can be used which is a computational efficient variant for computing the DFT \cite{Cooley1965}. Here, a frame is zero-padded to the next radix-2 number. In case of a radix-2 FFT implementation the amount of operations are $N_f/2\cdot log_2(N_f)$ complex multiplies and $N_f\cdot log_2(N_f)$ complex additions with $N_f$ the length of the (zero-padded) input frame. A complex multiplication is assumed to consist of 4 multiplications and 2 additions. By assuming an in-place algorithm, the total needed storage is $N_t\cdot S$ bits along with $5\cdot N_f\cdot log_2(N_f)$ memory accesses.

\subsubsection{log Mel-frequency transform}
As a first step only the power spectrum $|\textbf{z}|^2$ up to $N_f/2$ samples is retained since studies have shown that the amplitude of the spectrum is of more importance than the phase information. Then, the Mel-frequency filterbank smooths the high-dimensional magnitude spectrum such that it reflects the sensitivity of the human auditory system to frequency shifts where the lower frequencies are perceptually of more importance than the higher ones. The filterbank is defined by overlapping triangular frequency response bandpass filters with a constant spacing and bandwidth in the Mel-frequency domain. 


Typically, the number of bands $N_m$ is set in the range between 20 and 60 \cite{DCASE2013,DCASE2016}. The log Mel features can be computed by using:
\begin{equation}
\textbf{m}_b = \log{(\sum_{k=0}^{N_f/2-1}\textbf{W}_{bk}|\textbf{z}_k|^2)}
\label{eq:melTrans}
\enspace,
\end{equation}
with $b=[0,1,\dots,N_m-1]$ and $\textbf{W}_{bk}$ the weight of the Mel filter bank in band $b$ at frequency $k$. In a final step a logarithm operation is performed on the Mel features, which is also motivated by human perception of sound as humans hear loudness on a logarithmic scale. The amount of operations summarises to $N_f/2\cdot N_m$ multiply-accumulates and $N_m$ logarithms. In total $(N_f/2\cdot N_m + N_m)\cdot S$ bits need to be stored along with $2\cdot N_f\cdot N_m  + 2\cdot N_m$ memory accesses.

\subsubsection{Discrete Cosine Transform}
The Discrete Cosine Transform (DCT) expresses the Mel features in terms of a sum of cosine functions. These cosine functions describe the amplitude envelope of the Mel features. A limited set of coefficients (typically 14) are retained as they contain the elementary aspects of the shape. The DCT on the log Mel features is defined as:
\begin{equation}
\textbf{c}_d =\sum_{b=0}^{N_m-1}\textbf{m}_b\cos\Big[d\Big(b+\frac{1}{2}\Big)\frac{\pi}{N_m}\Big]
\enspace,
\label{eq:dctTrans}
\end{equation}
with $d=[0,1,\dots,D-1]$ and $[\textbf{c}_0, \textbf{c}_1, \dots, \textbf{c}_{N_c-1}]$ as the MFCC feature vector with length $N_c$. The amount of operations consist of $N_m\cdot N_c$ multiply-accumulates.  In total $(N_m + 1)\cdot N_c$ bits need to be stored along with $4\cdot N_m\cdot N_c$ memory accesses.

%

\subsection{Classifier: Artificial Neural Network}

Artificial Neural Networks, inspired by biological neural networks, automatically learn tasks based on provided data (and desired output) \cite{Sze2017}. A (Deep) Neural Network architecture is highly customizable and constructed using multiple layers. Following subsections roughly introduce several common layers used in (Deep) Neural Network architectures along with the amount of operations. 

\subsubsection{Fully-Connected layer}

The main building-block of the Fully Connected (FC) layer is an artificial neuron which is modelled using a modified version of a perceptron. A perceptron is a linear classifier able to discriminate two classes \cite{Rosenblatt1958}. The formal definition of the modified perceptron is $f(x) = \sigma(\textbf{w}^{T} \textbf{x})$, with $\textbf{x}$ the input vector, $\textbf{w}$ the learned weight vector and $\sigma$ an activation function to (non-linearly) transform the output value. The input vector $\textbf{x}$ is augmented with a scalar of value $1$ to allow for a shift in the discriminating hyperplane. Different from the original perceptron is that the activation function is not restricted to a threshold. A perceptron can be extended to multi-class classification by stacking multiple perceptrons (one for each class) and is referred as a multi-class perceptron. When used in Neural Networks, this is denoted as a FC layer. To allow for non-linear classification multiple FC layers can be concatenated where each output of the previous layer is connected to all inputs of that particular layer. In between those layers, a non-linear activation function should be used to create the non-linearity. As output layer the activation function typically consists of a Softmax operation to provide a probabilistic output. These activations are defined in section \ref{ss:act}. The amount of operations for the FC layer are $L_n(L_i+1)$ multiply-accumulates, with $L_n$ the amount of neurons and $L_i$ the size of the input vector. In total $L_n\cdot (L_i+2)\cdot S$ bits need to be stored along with $4\cdot L_n\cdot (L_i+1)$ memory accesses.

\subsubsection{Activation function}
\label{ss:act}

Activation functions introduce non-linear properties to the Neural Network. It does a non-linear mapping on the output of a artificial neuron to either provide input to a following layer or a probabilistic output on the final layer.  Numerous activations have been proposed over past decades. Currently, the most commonly used are the Rectified Linear Unit (ReLU), Logistic and Tanh \cite{DCASE2017}. ReLU is described as $\sigma(\textbf{z}_k) = \max{(0,\textbf{z}_k)}$ with $\textbf{z}$ the output of the previous layer and $k=[0,1,\dots,L_n-1]$. As it does not provide a probabilistic output, it is only used in-between layers. Logistic and Tanh are defined as $\sigma(\textbf{z}_k) = {1}/{(1+e^{-\textbf{z}_k})}$ and $\sigma(\textbf{z}_k) = {2}/{(1+e^{-2\textbf{z}_k})}-1$ respectively. As output layer, a Softmax function is typically used to compute the probabilities for each class. The Softmax is defined as $\sigma(\textbf{z}_k) = {e^{\textbf{z}_k}}/{\sum_{k=0}^{K-1}e^{\textbf{z}_k}}$. For ReLU, the total amount of operations devoted to the activation function is $L_n$ comparisons along with $3\cdot L_n$ memory accesses. The Tanh activation function consists of $2\cdot L_n$ additions, and $L_n$ divisions and exponentials along with $12\cdot L_n$ memory accesses. In case of Softmax and Logistic the total amount of operations summarises to $L_n$ additions, divisions and exponentials along with $9\cdot L_n$ memory accesses.

\subsubsection{Convolutional layer}

In case of more-than-one-dimensional input data, connecting all inputs to a FC layer might lead to an unreasonable amount of weights. A convolutional layer is similar to a FC layer as they are also made up of artificial neurons that have learnable weights. A convolution layer however, convolves the input data with multiple so-called templates. It is assumed that a particular template, smaller than the input data, can be reused at multiple positions in the input data. At each convolution index these templates are locally-connected to the input data and output one activation. Due to the weight sharing the amount of weights is reduced compared to directly using a FC layer. 

The hyperparameters that define a convolutional layer are the number of templates $T_{n}$, template dimensions $T_{d,k}$, convolution strides $T_{s,k}$ and amount of zero padding $T_{p,k}$ on the input data for a particular dimension index $k$. The output size of a convolution layer for a particular template $T_n$ is defined as $L_{o,k} = (L_{i,k} - T_{d,k} +  2T_{p,k})/T_{s,k}+1$, with $L_{i,k}$ and $L_{o,k}$ the length of the input- and output data respectively at dimension $k$. This leads to total amount of operations of $T_n\cdot\prod_{k=0}^{L_d}((L_{i,k} - T_{d,k} +  2T_{p,k})/T_{s,k}+1)\cdot(\prod_{k=0}^{L_d}T_{d,k}+1)$ multiply-accumulates. The amount of memory accesses are the amount of operations multiplied by 4. The total needed storage consists of $T_n\cdot(\prod_{k=0}^{L_d}T_{d,k} + 1)\cdot S$ bits for the weights and $(\prod_{k=0}^{L_d}L_{o,k})\cdot S$ bits as output.

\subsubsection{Pooling layer}

It is a common practice to introduce a pooling layer in-between succesive convolutional layers. Such a layer performs undersampling on the previous output to reduce overfitting and the amount of parameters and computation in the network. Similar to a convolutional layer, the pooling layer iterates over the entire input data. Different from a convolutional layer is that, instead of a matrix product of the locally-connected data and a template, it calculates a summary of the current locally-connected part of the data. This summary could either be average or a max. operation. The pooling layer therefor consists of $\prod_{k=0}^{L_d}((L_{i,k} - T_{d,k} +  2T_{p,k})/T_{s,k}+1)\cdot(\prod_{k=0}^{L_d}T_{d,k}-1)$ operations.  The amount of memory accesses are the amount of operations multiplied by 3. The needed output storage summarises to $(\prod_{k=0}^{L_d}L_{o,k})\cdot S$ bits. 

\subsubsection{Batch Normalization}

Batch Normalization (BN) was introduced to compensate for the so-called \textit{internal covariate shift} that slows down the training of the network \cite{Ioffe2015}. BN performs a standard normalization, during training on each mini-batch, on the output of the activations of each layer seperately. During the test phase this adds an additional shift and scale to each activation output. For a particular layer the amount of operations are $L_n$ additions and multiplications, which results in $6\cdot L_n$ memory accesses. The total stored information is $2\cdot L_n \cdot S$ bits for the shift and scale.

%% file: text/wirelessconsumption.tex
\section{Energy consumption of communications}
\label{sec:comm}

This section focuses in describing and modeling the energy consumption of the communication module of the sensor node. We assume that the node is equipped with $N_t$ transmitter and $N_r$ receiveing  antennas and corresponding transceiver branches, respectively~\cite{rosas2015impact}. If a node has only one antenna then $N_t=N_r=1$ is used. 

By default, the node is assumed to be in a low power consumption (sleep) mode~\cite{karl2007protocols}. At its designated time the node wakes up, and engages a transmission and reception of frames with the central connection point respectively. In case of transmission, if $x$ attempts are decoded with errors, the transmitter will declare an outage and go to sleep mode for one coherence time of the channel. Let us denote as $\tau_\text{out}$ the number of outages and $\tau_x$ the number of transmission trials required to achieve a decoding without errors ($\tau_x \in \{1,\dots,x\}$). These are random variables with mean values given by $\bar{\tau}_\text{out}$ and $\bar{\tau}_x$, whose values depends on the modulation, coding scheme and fading statistics~\cite{rosas2012modulation}.

The energy consumption per correctly transmitted information bit, $\energiabar{T}$, can be modeled as~\cite{rosas2016optimizing}
\begin{equation} \label{e:ET}
\bar{\mathcal{E}}_\text{T} = (1 + \bar{\tau}_\text{out}) \frac{\energia{st}}{N_T} + \energia{enc} + \left( \energia{etx,b} + \energia{PA,b} + \energia{erx,fb}\right) ( x \bar{\tau}_\text{out} + \bar{\tau}_x)
\enspace.
\end{equation}
Above, $\energia{st}$ is the startup energy required to awake the node from the low power mode, $\energia{enc}$ is the energy required to encode the forward message, $\energia{etx,b}$ and $\energia{erx,fb}$ are the energy consumption of the baseband and radio-frequency electronic components that perform the forward transmission and the reception of the feedback frame respectively and $\energia{PA,b}$ is the energy consumption of the power amplifier (which is responsible of the electromagnetic irradiation) for sending an information bit.

By analogy, the total energy used per correctly received bit, which involves demodulating forward frames and transmitting the feedback frames, can be modeled as \cite{rosas2016optimizing}
\begin{equation} \label{e:ER}
\bar{\mathcal{E}}_\text{R} = (1 + \bar{\tau}_\text{out}) \frac{\energia{st}}{N_R} + \left(\energia{dec} + \energia{erx,b} + \energia{etx,fb} +  \energia{PA,fb} \right) ( x \bar{\tau}_\text{out} + \bar{\tau}_x)
\enspace.
\end{equation}
Above, $\energia{erx,b}$ and $\energia{etx,fb}$ are the energy consumption of the baseband and radio-frequency electronic components that perform the forward reception and the transmission of the feedback frame respectively and $\energia{PA,fb}$ is the energy consumption of the power amplifier for transmitting feedback frames.

\subsection{Modeling the energy consumption of the PA} 

Let us express $\energia{PA}$, the total consumption due to the irradiated power, as
\begin{equation}\label{eq:PA00}
\energia{PA} = \energia{PA,b} + \energia{PA,fb} = \sum_{j=1}^{\Nt} \P{PA}^{(j)} \T{b} +  \sum_{j=1}^{\Nt} \P{PA}^{(j)} \T{fb}
\enspace,
\end{equation}
where $\P{PA}^{(j)}$ is the power consumption of the PA of the $j$-th transceiver branch. The total time per information bit in the forward direction, $T_\text{b}$, is calculated as~\cite{rosas2015impact}
\begin{equation}
  \label{tiempo_de_bit}
  \T{b} = \frac{1}{rR_\text{s}} \left( \frac{1}{\omega b} + \frac{H}{\omega L}+ \frac{N_\text{t} O_\text{a}+O_\text{b}}{L} \right)
\enspace,
\end{equation}
where $R_\text{s}$ is the physical layer symbol-rate, $r$ is the code rate of the coding scheme (percentage of data per payload bits), $\omega$ is the multiplexing gain of the MIMO modulation, $b=\log_2( M)$ is the number of bits per complex symbol, $H$ and $L$ are the number of bits in the header and payload of the frame, $O_\text{a}$ is the acquisition overhead per transceiver branch and $O_\text{b}$ is the remaining overhead, which is approximately independent of the antenna array size (both $O_\text{a}$ and  $O_\text{b}$ are measured in bits~\cite{cui2005}). Similarly, the total time per feedback frame, $T_\text{fb}$, is given by
\begin{equation}
  \label{tiempo_de_fbbit}
    \T{fb}=\frac{F}{r\omega R_\text{s}L}
\enspace,
\end{equation}
where $F$ is the number of bits of the feedback frame.

Let us relate the power consumption of the PAs with the signal-to-noise ratio (SNR). The $j$-th transmit antenna radiates $\P{tx}^{(j)}$ watts, which are provided by the corresponding power amplifier (PA). The PA's power consumption is modeled by~\cite{cui2005}
\begin{equation} \label{e:Ppa}
\P{PA}^{(j)} = \frac{1}{\eta} \P{tx}^{(j)}
\enspace,
\end{equation}
where $\eta$ the average efficiency of the PA. In general, the average PA efficiency can be more precisely modeled using the distribution of the output power of the underlying signal. If we limit the analysis to linear PAs, such as Class A and Class B PAs (as many mobile and wireless communication devices require linear PAs), then we can approximate $\eta$ with 
\begin{equation} \label{e:Ppa}
\eta = \left( \frac{\Pbar{tx}}{P_\text{max}} \right)^{\beta}\eta_\text{max}
\enspace,
\end{equation}
where $\Pbar{tx}$ is the average radiated power (which we assume is the same for all transmitter antennas), $P_\text{max}$ is the maximal PA output and 
\begin{align} \label{e:Ppa}
\eta_\text{max, class A} = 0.5 \quad &\text{and} \quad \beta_\text{class A} = 1 \\
\eta_\text{max, class B} = 0.785 \quad &\text{and} \quad \beta_\text{class B} = 0.5
\enspace.
\end{align}
In these equations, $\P{back-off} = \P{max} / \Pbar{tx}$ is the back-off of the PA. Highest efficiency is achieved by constant-envelope signals for which $\P{back-off}=1$. In general, one can calculate the back-off coefficient as $\P{back-off} = \xi/S$, where $\xi$ is the peak-to-average power ratio of the modulation (which is usually calculated as $\xi = 3(\sqrt{M}-1)/(\sqrt{m}+1)$) and $S$ accounts any additional back-off that may be taken when the wireless link has excess link budget and transmit power can be decreased further. Finally, the relationship between the PA consumption $P_\text{PA}$ and the radiated power $\P{tx}$ is calculated as:
\begin{equation} \label{e:Ppa}
\P{tx}^{(j)} = \left( \frac{S}{\xi} \right)^\beta \eta_\text{max} \P{PA}^{(j)}
\enspace.
\end{equation}

The transmission power attenuates over the air with path loss and arrives at the receiver with a mean power given by
\begin{equation} \label{e:Prx} 
P_{\mathrm{rx}}^{(j)} = \frac{P_{\mathrm{tx}}^{(j)}}{A_0d^{\alpha}} 
\enspace,
\end{equation}
where $d$ is the distance between transmitter and receiver and $\alpha$ is the path loss exponent. Above, $A_0$ is a parameter that is defined by the free-space friss equation
\begin{equation}
A_0 = \frac{1}{G_\text{t} G_\text{r}} \left(\frac{4\pi}{\lambda}\right)^2
\enspace,
\end{equation}
where $G_\text{t}$ and $G_\text{r}$ are the transmitter and receiver antenna gains and $\lambda$ is the carrier wavelength. Finally, if $\sigma_\text{s}^2$ is the average received power per symbol at the input point of the decision stage of the receiver (which is located after the MIMO decoder), the total received signal power is given by
\begin{equation}\label{ooSS}
\sum_{j=1}^{\Nt}   \bar{P}_\text{rx}^{(j)} = \omega \sigma_\text{s}^2 = \omega \sigma_\text{n}^2 \bar{\gamma}
\enspace,
\end{equation}
where $\sigma_\text{n}^2$ is the thermal noise power and $\bar{\gamma}$ is the average SNR. In general, $\sigma_\text{n}^2=N_0WN_\text{f}M_\text{L}$, where $N_0$ is the power spectral density of the baseband-equivalent additive white Gaussian noise (AWGN), $W$ is the transmission bandwidth, $N_\text{f}$ is the noise figure of the receiver's front end and $M_\text{L}$ is a link margin term which represents any other additive noise or interference \cite{cui2004}.

With all this, one finds that
\begin{align}
\sum_{j=1}^{\Nt} \P{PA}^{(j)} &=  \left( \frac{\xi}{S} \right)^\beta \frac{1}{\eta_\text{max}}\sum_{j=1}^{\Nt} \P{tx}^{(j)} \\
&= \left( \frac{\xi }{S} \right)^\beta \frac{A_0d^{\alpha}}{\eta_\text{max}} \sum_{j=1}^{\Nt} P_{\mathrm{rx}}^{(j)} \\
&= \left( \frac{\xi }{S } \right)^\beta \frac{N_0WN_\text{f}M_\text{L}  A_0 }{\eta_\text{max} } \omega d^{\alpha} \bar{\gamma}\\
&= A \omega d^\alpha \bar{\gamma}
\end{align}
with $A$ a constant.

\subsection{Modeling the energy consumption of the other electronic components}

\begin{figure}[H]
	\centering
	\includegraphics[width=1\textwidth]{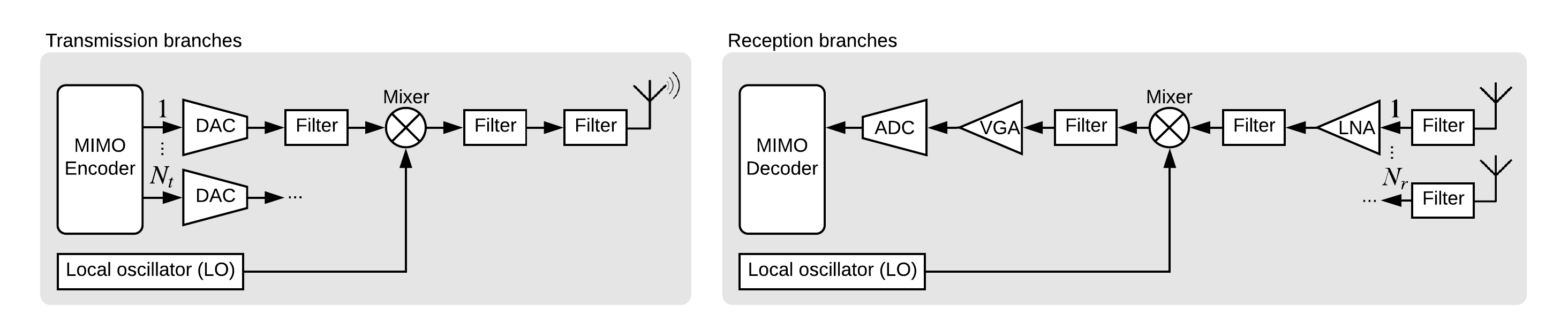}
	\caption{MIMO architecture considered in this work}
	\label{fig:MIMOschematics}
\end{figure}

Let us assume that the device is equipped with $\Nt$ antennas, using an architecture as shown in Figure~\ref{fig:MIMOschematics}. Then, one can express the electronic consumption of the transmitter per information bit as~\cite{rosas2015impact}
\begin{equation}
\energia{etx,b} = \Nt \P{etx}\T{b} = \Nt ( \P{DAC} + 2\P{filter} + \P{LO} + \P{mixer})\T{b}
\enspace,
\end{equation}
where $P_\text{etx}$ is the power consumption of the electronic components (filters, mixer, DAC and local oscillator) that perform the transmission per transceiver branch. A similar equation for $\energia{etx,fb}$ can be obtained by replacing $T_\text{b}$  with $T_\text{fb}$ which are defined in \eqref{tiempo_de_bit} and \eqref{tiempo_de_fbbit} respectively.

Analogously, one can show that
\begin{align}
\energia{erx,b} = \Nr\P{erx}\T{b}= \Nr( 3 \P{filter} + \P{LNA} + \P{LO} + \P{mixer} + \P{VGA} + \P{ADC})\T{b}
\enspace,
\end{align}
where $\P{erx}$ is the power consumption of the electronic components (filters, mixer, ADC, VGA and local oscillator) that perform the forward and feedback frame reception per branch. A similar equation can be obtained for $\energia{erx,fb}$.

Following \cite{cui2005}, we will model the energy consumption of DACs as $\P{DAC} = \beta (\P{DAC}^\text{static} + \P{DAC}^\text{dyn})$, where $\P{DAC}^\text{static}$ (resp. $\P{DAC}^\text{static}$) is the static (resp. dynamic) power consumption and $\beta$ is a correcting factor to incorporate some second order effects. If a binary-weighted current-steering DAC is considered \cite{Gustavsson2000}, then
\begin{equation}
\P{DAC}^\text{static} = V_\text{dd} I_\text{unit} \E{ \sum_{i=0}^{n_1} 2^ib_i} = \frac{1}{2} V_\text{dd} I_\text{unit} (2^{n_1} -1)
\enspace,
\end{equation} 
where $n_1$ is the resolution, $b_i$ are independent Bernoulli random variables with parameter $1/2$, $V_\text{dd}$ is the power supply voltage and $I_\text{unit}$ is the unit current source corresponding to the least significant bit. The dynamic consumption can be approximated as $\P{DAC}^\text{dyn} = (1/2)n_1 C_\text{p}f_\text{s}^\text{DAC} V_\text{dd}^2$, where $C_\text{p}$ is the parasitic capacitance of each switch, the $1/2$ is the switching probability and $f_\text{s}^\text{DAC}$ is the sampling frequency. Hence, the total consumption of the DAC is expressed as
\begin{equation}
\P{DAC} = \frac{\beta}{2} \left[ V_\text{dd}I_\text{unit}(2^{n_1}-1) + n_1C_\text{p} f_\text{s}^\text{DAC}V_\text{dd}^2\right]
\enspace.
\end{equation}
In turn, the ADC consumption can be computed using \eqref{eq:ADCconsumption}.


\subsection{Modeling the energy consumption of encoding and decoding forward frames}
\label{s:encoding_decoding} 
The computations requred for encoding and decoding data frames can be demanding, depending on the choice of coding scheme~\cite{rosas2016optimizing}. Therefore, it makes sense to include the energy costs of these operations into the energy budget. However, for simplicity we neglect the coding and decoding costs of headers and feedback frames, which usually are either uncoded or use lightweigth codes whose processing can safely be neglected.

By considering that the encoding has to be done for each frame, its cost is shared among the $L_u$ payload bits. Similarly to \eqref{eq:mPconsumption}, the energy consumption for encoding one frame --- normalized per payload bit --- is given by
\begin{equation}
\label{eq:enc}
\energia{enc} = \frac{1}{N_T} \energia{cc} \sum_{j=1}^{J_\text{ALU}}  c_j n_j^\text{enc}
\enspace,
\end{equation}
with $n_j^{enc}$ the amount of times the $j$-th arithmetic operation is performed. Note that it is straightforward to write an equation for the decoding cost equivalent to \eqref{eq:enc}. More information on the computational complexity of the used error correction code (BCH code) can be retrieved in \cite{rosas2016optimizing}.

Finally, following \cite{rosas2016optimizing} our modeling does not includes the cost of memory storage and access, which is left for future work.

\subsection{Re-transmission Statistics}
\label{s:re-transmission_stats}

For computing the statistic of retransmissions due to decoding errors, let us derive an expression for $\bar{\tau}_\text{out}$ and $\bar{\tau}_x$ following~\cite{rosas2016optimizing}. As each outage declaration are independent events, $\tau_\text{out}$ will be a geometric random variable with p.d.f. $\mathbb{P}\{\tau_\text{out} = j\} = (1-q_x)q_x^j$, where $q_x= 1 - \mathbb{P}\{\tau \leq x\}$ is the outage probability. Then, a direct calculation shows that its mean value is given by
\begin{equation}\label{eq:mean_outage}
 \bar{\tau}_\text{out}=\frac{q_x}{1-q_x}
\enspace.
\end{equation}
The p.d.f. of $\tau_x$ is given by
\begin{equation}
\mathbb{P}\{\tau_x =t\}= \mathbb{P}\{ \tau = t \vert \tau \leq x \} = \frac{ \mathbb{P}\{ \tau = t\}}{1- q_x }
\end{equation}
and, hence, it mean value is calculated as
\begin{equation}\label{eq:tauX}
\bar{\tau}_x = \sum_{t=1}^x t \cdot\mathbb{P}\{ \tau_x = t \} = \frac{1}{1-q_x}\sum_{t=1}^x t \cdot\mathbb{P}\{ \tau = t \}
\enspace.
\end{equation}
Finally, one can find that 
\begin{equation}\label{eq:tau_eff}
x\bar{\tau}_\text{out} + \bar{\tau}_x = \frac{1}{1-q_x}\left( x q_x + \sum_{t=1}^x t \mathbb{P}\{\tau=t\} \right)
\enspace.
\end{equation}

The values of $\bar{\tau}_\text{out}$ and $\bar{\tau}_x$ depend strongly on the correlations of the wireless channels. Let us consider two extreme examples: fast fading and static fading channels. In fast fading channels the frame error rate of each transmission trial are i.i.d. random variables, while in static channels there are the same random variable. Then, one have that $q_x^{\text{ff}} = \bar{ P}_\text{f}^x$ and $q_x^{\text{bf}} =  \E{ P_\text{f}^x}$, and a direct application of the Jensen inequality shows that $q_\text{x}^\text{ff} \leq q_\text{x}^\text{bf}$. Also, by defining $\Phi(x)$ as
\begin{equation}
\Phi(x) = \frac{1}{1 - \E{ P_\text{f}^x}}  \E{ \frac{1-P_\text{f}^x}{1-P_\text{f}}} \label{eq:tau_bf}
\enspace,
\end{equation}
it can be shown that $x\bar{\tau}_x^\text{bf} + \bar{\tau}_\text{out}^\text{bf}=\Phi(x)$ and $x\bar{\tau}_x^\text{ff} + \bar{\tau}_\text{out}^\text{ff}=\Phi(1)$. It can also be shown that $\Phi(x)$ is a increasing function, so that in fast fading scenarios one usually requires more transmissions trails.

Particular functional forms can be given for $P_\text{f}$ depending on the error correcting code scheme in use. For the sake of concreteness, in the sequel we present a derivation valid for BCH codes, following the derivation shown in ~\cite{rosas2014optimizing}. Let us denote as $n$ the legth of each codeword, and assuming that $n < L$ let's define $\nc=L/n$ ($\nc \in \mathbb{N}$) as the number of codewords per payload. In order to decode a frame correctly, one needs to correctly obtain $H$ correct header symbols and $\nc$ codewords with at least $(n - t)=\lambda$ correct symbols, where $t$ is the maximum number of bits that the FEC block code is able to correct in each codeword. Therefore, by taking into consideration the various possible permutations, $\bar{P}_\text{f}$ can be expressed in terms of the bit error rate of the $M$-ary modulation $P_\mathrm{b}(\gamma)$ and the binary modulation symbol error rate $P_\text{bin}(\gamma)$ as
\begin{align}
\label{eq:SER3_nc}
\meanP{f}(\meang) =  1 - \left[ 1 - \bar{P}_\text{bin}(\bar{\gamma})  \right]^{H} \left[ \sum_{j=0}^t  \binom{n}{j} \left[ 1 -  \bar{P}_\text{b}(\bar{\gamma}) \right]^{n-j} \bar{P}_\text{b}(\bar{\gamma})^j \right]^{n_c}
\enspace.
\end{align}
Above, $\gamma$ corresponds to the signal-to-noise ratio and $\bar{\gamma}=\E{\gamma}$. Also, please note that we are using the shorthand notation  $\bar{P}_{\text{bin}}(\bar{\gamma})=\mathbb{E}\{P_{\text{bin}}(\gamma)\}$ and $\bar{P}_{\text{b}}(\bar{\gamma})=\mathbb{E}\{P_{\text{b}}(\gamma)\}$, and that \eqref{eq:SER3_nc} is only valid for scenarios that experience fast-fading conditions.

Finally, let us remark that simple methods to approximate the error rates of MIMO channels are available in~\cite{rosas2013nakagami}.

\subsection{Final model}

Using the material presented in the above subsections, one an finally express \eqref{e:ET} as
\begin{align} \label{e:ETfin}
\bar{\mathcal{E}}_\text{T} &=  \frac{1}{N_T}(\frac{\energia{st}}{1-q_x} + \energia{cc} \sum_{j=1}^{J_\text{ALU}}  c_j n_j^\text{enc}) + \left[ \left( \Nt \P{etx} + A d^\alpha \omega \bar{\gamma} \right) \T{b} + \Nt\P{erx} \T{fb} \right] \Phi(x)
\enspace.
\end{align}

This equation allows to express the average energy consumption per successfully transferred bit in terms of a number of design parameters. Similarly, the average energy consumption per succesfully received bit, given by \eqref{e:ER}, is expanded to

\begin{equation} \label{e:ER}
\bar{\mathcal{E}}_\text{R} = \frac{\energia{st}}{(1-q_x) N_R} + \left(\frac{1}{N_R}\energia{cc} \sum_{j=1}^{J_\text{ALU}}  c_j n_j^\text{dec} + \Nr\P{erx}\T{b}  + (\Nr \P{etx} +   A d^\alpha \omega \bar{\gamma})\T{fb} \right) \Phi(x)
\enspace.
\end{equation}

An overview of the relevant hardware related parameters along with the used values is given in Appendix \ref{appendix:params} Table \ref{tab:paramcom}.

%% file: text/modelanalysis.tex
\section{Model analysis}
\label{sec:modanalysis}

The proposed model contains separate blocks for each of the four considered layers (sensing, processing and communication), and each of them are populated by a considerable amount of parameters. However, some of these parameters are more interesting to be explored than others. Although we don't develop an exhaustive exploration of all these parameters, 
the remaining of this section provides some guidelines on aspects of interest that can be explored.

Table \ref{tab:expparam} provides a summary of some parameters from the sensing and communication layers that have a strong influence on the energy efficiency of the sensor node.\footnote{The processing layer is not covered in the table; one can compare various feature extraction and classifier architectures, or no processing at all.} In case of the sensing layer, $f_{s,mic}$ and $n_{mic}$ are interesting parameters as they have a big impact on the consumed energy 
because it regulate the amount of information to be processed and communicated. Naturally, more collected information equals a higher energy consumption.

In case of the communication layer, one parameter to explore is the amount of communicated bits $N_T$ that depends on both the processing and sensing layer. An interesting trade-off is the energy spend in processing versus the one spent in communication. Another parameter of interest is the number of bits $t$ the FEC can correct. A higher number leads to more communication frame overhead but less re-transmission due to errors, as shown in Refs.~\cite{rosas2014optimizing,rosas2016optimizing}. Another parameter of interest is the transmission bandwidth $W$, which is determined usually be the communication standard that the network uses (Zigbee, Bluetooth, etc). In general a higher bandwidth is beneficial as transmission times -- and hence the baseline electronic consumption of the transceiver -- are reduced. Please note that the effect of interference is not included in this model, being left for future work. Finally, one can explore the impact of various the communication channel via the path-loss coefficient $\alpha$ and the SNR $e$. In busy scenarios with many obstables the path-loss is higher and the received signal strength is smaller, which usually causes more re-transmission and, in turn, a lower energy efficiency. 

\begin{table}[h]
\begin{center}
\begin{tabular}{|c|c|c|} \hline \hline
Parameter                 		& Description                                                				& Value             
\\ \hline \hline
$f_{s,mic}$                         	& Sampling frequency of the microphone				& 16 kHz \\
$n_{mic}$                         		& ADC resolution								& 12 bit\\
\hline
$N_T$                         		& Amount of bits to communicate						& -\\
$t$   					& Number of bits the FEC can correct					& 4  \\
$W$   					& Transmission bandwidth							& 1 MHz \\
$\alpha$   				& Path-loss coefficient							& 3.2 \\
$e$   					& SNR of communication							& 25 dB \\
\hline\hline 
\end{tabular}
\caption{Experimental parameters along with the default value in the source code}
\label{tab:expparam}
\end{center}
\end{table}
%


%% file: text/appendix.tex

\appendix
\section{Default parameters used in the MATLAB implementation}
\label{appendix:params}

\begin{table}[h]
\begin{center}
\begin{tabular}{|c|c|c|} \hline \hline
Parameter                 			& Description                                                		& Value             \\ \hline \hline
$T$                         				& Room temperature                              			& 290 K \\
$\P{mic,act}$           			& Active microphone power consumption 		& 10 mW \\
$V_\text{dd}^\text{LNA,mic}$    	& Voltage supply of the LNA-mic 				&  1.5 V  \\  
$\text{NEF}$           			& Noise efficiency factor             				& 6  \\
$\text{FOM}_\text{ADC}$ 		& Figure of merit of the ADC  				& 500 fJ/conv. \cite{ADClink2014} \\ 
$f_{s,mic}$                         		& Sampling frequency of the microphone		& 16 kHz \\
$n_{mic}$                         			& ADC resolution						& 12 bit\\
\hline\hline 
\end{tabular}
\caption{Overview of the default parameters used in the MATLAB implementation for the sensing layer}
\label{tab:paramsens}
\end{center}
\end{table}

\begin{table}[h!]
\begin{center}
\begin{tabular}{|c|c|c|} \hline \hline
Parameter                 		& Description                                                		& Value             
\\ \hline \hline
$\energia{op}$   			& Energy per operation (GP proc.)    			& 500 pJ \cite{Horowitz14, CortexM4}  \\
$$   					& Energy per operation (GP DSP)    			& 100 pJ \cite{Horowitz14} \\
\hline 
$\energia{ma}$   			& Energy per memory access (on-chip SRAM) 		& 100 fJ/bit  \cite{Haine17}\\
$$   					& Energy per memory access (off-chip SRAM) 		& 100 pJ/bit \cite{CypressSRAM}\\
$$   					& Energy per memory access (off-chip DRAM) 		& 100 pJ/bit \cite{Horowitz14}\\
\hline 
$\energia{ms}$   			& Energy memory leakage (on-chip SRAM)		& 50 pW/bit  \cite{Haine17}\\
$$   					& Energy memory leakage (off-chip SRAM)		& 10 pW/bit \cite{CypressSRAM}\\
$$   					& Energy memory leakage (off-chip DRAM)		& 75 pW/bit \cite{Horowitz14}\\
\hline 
$S$          				& Word size							& 32 bit \\
\hline 
$c_\text{mac}$          		& Multiply-accumulate cost					& 2 ops. \cite{CortexM4}\\
\hline 
$c_\text{add}$         		& Addition cost           					& 1 op.  \cite{CortexM4}\\ 
\hline 
$c_\text{mul}$         		& Multiplication cost       					& 1 op. \cite{CortexM4}\\ 
\hline 
$c_\text{div}$         		& Division cost          					& 8 op.  \cite{CortexM4}\\ 
\hline 
$c_\text{cmp}$         		& Comparator cost						& 1 op.  \cite{CortexM4}\\ 
\hline 
$c_\text{exp}$         		& Natural exponential cost 				& 2 op.  \cite{CortexM4}\\ 
\hline 
$c_\text{log}$         		& Logarithmic cost      					& 25 ops.  \cite{CortexM4}\\ 
\hline\hline 
\end{tabular}
\caption{Overview of the default parameters used in the MATLAB implementation for the processing layer}
\label{tab:paramproc}
\end{center}
\end{table}
\begin{table}[h!]
\begin{center}
\begin{tabular}{|c|c|c|} \hline \hline
Parameter                 			& Description                                                		& Value             \\ \hline \hline
$\ene_\mathrm{st}$   			& Start-up energy                                          		& 94 $\mu$J \cite{Siekkinen2012}\\ 
$P_\text{filter}$      			& Filter power consumption          			& 1 mW \cite{Balankutty2010} \\ 
$P_\text{mixer}$     			& Mixer power consumption          			& 1 mW \cite{Balankutty2010} \\
$P_\text{LNA,rx}$      			& LNA power consumption          				& 3 mW \cite{Balankutty2010} \\
$P_\text{VGA}$      			& VGA power consumption          				& 5 mW \cite{Balankutty2010} \\
$\P{LO}$              				& Local oscilator consumption     				& 22.5 mW \cite{Balankutty2010} \\ 
\hline
$n_1$                             			& Resolution of the Tx DAC                 			& 10 levels \cite{cui2005}\\
$f_\text{s}^\text{DAC}$      		& DAC sampling frequency                  			& 4 MHz \\
$V_\text{dd}^\text{DAC}$   		& Voltage supply of the DAC               			& 3 V \cite{cui2005} \\  
$I_\text{unit}$                  		& DAC unit current source                  			& 10 $\mu$A \cite{cui2005} \\
$C_\text{p}$                    			& DAC parasitic capacitance                			& 1 pF \cite{cui2005} \\  
$n_2$                             			& Resolution of Rx ADC                      			& 10 levels \cite{cui2005}\\
$f_\text{s}^\text{ADC}$      		& DAC sampling frequency                  			& 4 MHz \\
\hline
$\eta_\text{max}$       			& PA efficiency (Class B)                              		& 0.785 \%     \\                
$ \beta$              				& Exponent for Class B PA                      		& 0.5 \\
$ S $                 				& Additional back-off coefficient                 		& 0 dB \\ 
$G_\text{t}  $             			& Transmitter antenna gain                          		& 1.8 \\
$G_\text{r}  $             			& Receiver antenna gain                               		& 1.8 \\
\hline
$f_\text{c} $               			& Carrier frequency                                    		& 2.4 GHz \cite{IEEE-802.15.4short}\\ 
$W$                           			& Bandwidth                                                  		& 1 MHz \cite{Balankutty2010}    \\ 
$R_\mathrm{s} $        			& Symbol rate                                                		& 0.125 MBaud \cite{Balankutty2010} \\ 
$M$                        				& M-ary number                                             		& 2 (BPSK) \\     
\hline
$N_\mathrm{f}$         			& Receiver noise figure                               		& 16 dB \cite{Balankutty2010} \\
$M_\mathrm{l}$          			& Link margin                                              		& 20 dB     \\ 
$\alpha$                   			& Path-loss coefficient                                  		& 3.2 \\ 
\hline
$t$   						& Number of bits the FEC can correct			& 4  \\
\hline
$H$                          			& Frame Header                                            		& 2 bytes \cite{IEEE-802.15.4short} \\
$L$                           			& Payload                                                       		& 127 bytes \cite{IEEE-802.15.4short} \\
$O_\text{a}$                 			& Acquisition overhead                                   		& 4 bytes \cite{IEEE-802.15.4short} \\
$O_\text{b}$                			& Estimation and synchronization overhead      	& 1 bytes \cite{IEEE-802.15.4short} \\
$F$                            			& Feedback frame length                              		& 5 bytes \\ 
\hline\hline 
\end{tabular}
\caption{Overview of the default parameters used in the MATLAB implementation for the communication layer}
\label{tab:paramcom}
\end{center}
\end{table}


%% file: doc.bbl
\begin{thebibliography}{10}
\providecommand{\url}[1]{#1}
\csname url@samestyle\endcsname
\providecommand{\newblock}{\relax}
\providecommand{\bibinfo}[2]{#2}
\providecommand{\BIBentrySTDinterwordspacing}{\spaceskip=0pt\relax}
\providecommand{\BIBentryALTinterwordstretchfactor}{4}
\providecommand{\BIBentryALTinterwordspacing}{\spaceskip=\fontdimen2\font plus
\BIBentryALTinterwordstretchfactor\fontdimen3\font minus
  \fontdimen4\font\relax}
\providecommand{\BIBforeignlanguage}[2]{{%
\expandafter\ifx\csname l@#1\endcsname\relax
\typeout{** WARNING: IEEEtran.bst: No hyphenation pattern has been}%
\typeout{** loaded for the language `#1'. Using the pattern for}%
\typeout{** the default language instead.}%
\else
\language=\csname l@#1\endcsname
\fi
#2}}
\providecommand{\BIBdecl}{\relax}
\BIBdecl

\bibitem{Borkar2011}
S.~Borkar and A.~A. Chien, ``The future of microprocessors,'' \emph{Commun.
  ACM}, vol.~54, no.~5, pp. 67--77, May 2011.

\bibitem{Rawat2014}
P.~Rawat, K.~D. Singh, and J.~M. Chaouchi, Hakimaand~Bonnin, ``Wireless sensor
  networks: a survey on recent developments and potential synergies,''
  \emph{The Journal of Supercomputing}, vol.~68, no.~1, pp. 1--48, Apr 2014.

\bibitem{mainwaring2002wireless}
A.~Mainwaring, D.~Culler, J.~Polastre, R.~Szewczyk, and J.~Anderson, ``Wireless
  sensor networks for habitat monitoring,'' in \emph{Proceedings of the 1st ACM
  international workshop on Wireless sensor networks and applications}.\hskip
  1em plus 0.5em minus 0.4em\relax Acm, 2002, pp. 88--97.

\bibitem{Butun2014}
I.~Butun, S.~D. Morgera, and R.~Sankar, ``A survey of intrusion detection
  systems in wireless sensor networks,'' \emph{IEEE Communications Surveys
  Tutorials}, vol.~16, no.~1, pp. 266--282, First 2014.

\bibitem{Erden2016}
F.~Erden, S.~Velipasalar, A.~Z. Alkar, and A.~E. Cetin, ``Sensors in assisted
  living: A survey of signal and image processing methods,'' \emph{IEEE Signal
  Processing Magazine}, vol.~33, no.~2, pp. 36--44, March 2016.

\bibitem{Hashem2016}
I.~A.~T. Hashem, V.~Chang, N.~B. Anuar, K.~Adewole, I.~Yaqoob, A.~Gani,
  E.~Ahmed, and H.~Chiroma, ``The role of big data in smart city,''
  \emph{International Journal of Information Management}, vol.~36, no.~5, pp.
  748 -- 758, 2016.

\bibitem{Vacher2011A}
M.~Vacher, F.~Portet, A.~Fleury, and N.~Noury, ``Development of audio sensing
  technology for ambient assisted living: Applications and challenges,''
  \emph{International Journal of E-Health and Medical Communications (IJEHMC)},
  vol.~2, no.~1, pp. 35--54, January 2011.

\bibitem{Bertrand2011}
A.~Bertrand, ``Applications and trends in wireless acoustic sensor networks: A
  signal processing perspective,'' in \emph{2011 18th IEEE Symposium on
  Communications and Vehicular Technology in the Benelux (SCVT)}, Nov 2011, pp.
  1--6.

\bibitem{Lauwereins2018}
S.~Lauwereins, ``Cross-layer self-adaptivity for ultra-low power responsive iot
  devices,'' 2018.

\bibitem{karl2007protocols}
H.~Karl and A.~Willig, \emph{Protocols and architectures for wireless sensor
  networks}.\hskip 1em plus 0.5em minus 0.4em\relax John Wiley \& Sons, 2007.

\bibitem{anastasi2009energy}
G.~Anastasi, M.~Conti, M.~Di~Francesco, and A.~Passarella, ``Energy
  conservation in wireless sensor networks: A survey,'' \emph{Ad hoc networks},
  vol.~7, no.~3, pp. 537--568, 2009.

\bibitem{shih2001physical}
E.~Shih, S.-H. Cho, N.~Ickes, R.~Min, A.~Sinha, A.~Wang, and A.~Chandrakasan,
  ``Physical layer driven protocol and algorithm design for energy-efficient
  wireless sensor networks,'' in \emph{Proceedings of the 7th annual
  international conference on Mobile computing and networking}.\hskip 1em plus
  0.5em minus 0.4em\relax ACM, 2001, pp. 272--287.

\bibitem{ganesan2001highly}
D.~Ganesan, R.~Govindan, S.~Shenker, and D.~Estrin, ``Highly-resilient,
  energy-efficient multipath routing in wireless sensor networks,'' \emph{ACM
  SIGMOBILE Mobile Computing and Communications Review}, vol.~5, no.~4, pp.
  11--25, 2001.

\bibitem{rosas2015energy}
F.~Rosas, R.~D. Souza, M.~Verhelst, and S.~Pollin, ``Energy-efficient mimo
  multihop communications using the antenna selection scheme,'' in
  \emph{Wireless Communication Systems (ISWCS), 2015 International Symposium
  on}.\hskip 1em plus 0.5em minus 0.4em\relax IEEE, 2015, pp. 686--690.

\bibitem{ye2002energy}
W.~Ye, J.~Heidemann, and D.~Estrin, ``An energy-efficient mac protocol for
  wireless sensor networks,'' in \emph{INFOCOM 2002. Twenty-First Annual Joint
  Conference of the IEEE Computer and Communications Societies. Proceedings.
  IEEE}, vol.~3.\hskip 1em plus 0.5em minus 0.4em\relax IEEE, 2002, pp.
  1567--1576.

\bibitem{van2003adaptive}
T.~Van~Dam and K.~Langendoen, ``An adaptive energy-efficient mac protocol for
  wireless sensor networks,'' in \emph{Proceedings of the 1st international
  conference on Embedded networked sensor systems}.\hskip 1em plus 0.5em minus
  0.4em\relax ACM, 2003, pp. 171--180.

\bibitem{DCASE2013}
D.~Stowell, D.~Giannoulis, E.~Benetos, M.~Lagrange, and M.~D. Plumbley,
  ``Detection and classification of acoustic scenes and events,'' \emph{IEEE
  Transactions on Multimedia}, vol.~17, no.~10, pp. 1733--1746, Oct 2015.

\bibitem{DCASE2016}
A.~Mesaros, T.~Heittola, E.~Benetos, P.~Foster, M.~Lagrange, T.~Virtanen, and
  M.~D. Plumbley, ``Detection and classification of acoustic scenes and events:
  Outcome of the dcase 2016 challenge,'' \emph{IEEE/ACM Trans. Audio, Speech
  and Lang. Proc.}, vol.~26, no.~2, pp. 379--393, Feb. 2018.

\bibitem{DCASE2017}
A.~Mesaros, T.~Heittola, A.~Diment, B.~Elizalde, A.~Shah, E.~Vincent, B.~Raj,
  and T.~Virtanen, ``{DCASE 2017 Challenge setup: Tasks, datasets and baseline
  system},'' in \emph{Proceedings of the Detection and Classification of
  Acoustic Scenes and Events 2017 Workshop (DCASE2017)}, Munich, Germany,
  November 2017.

\bibitem{Yang2017}
T.~Yang, Y.~Chen, J.~Emer, and V.~Sze, ``A method to estimate the energy
  consumption of deep neural networks,'' in \emph{2017 51st Asilomar Conference
  on Signals, Systems, and Computers}, Oct 2017, pp. 1916--1920.

\bibitem{Chen2016}
Y.~Chen, J.~Emer, and V.~Sze, ``Eyeriss: A spatial architecture for
  energy-efficient dataflow for convolutional neural networks,'' in \emph{2016
  ACM/IEEE 43rd Annual International Symposium on Computer Architecture
  (ISCA)}, June 2016, pp. 367--379.

\bibitem{Steyaert1987}
M.~Steyaert and W.~Sansen, ``A micropower low-noise monolithic instrumentation
  amplifier for medical purposes,'' \emph{Solid-State Circuits, IEEE Journal
  of}, vol.~22, no.~6, pp. 1163--1168, Dec 1987.

\bibitem{ADClink2014}
\BIBentryALTinterwordspacing
B.~Murmann. Adc performance survey 1997-2014. [Online]. Available:
  \url{{http://www.stanford.edu/~murmann/adcsurvey.html}}
\BIBentrySTDinterwordspacing

\bibitem{Horowitz14}
M.~Horowitz, ``1.1 computing's energy problem (and what we can do about it),''
  in \emph{2014 IEEE International Solid-State Circuits Conference Digest of
  Technical Papers (ISSCC)}, Feb 2014, pp. 10--14.

\bibitem{Sze2017}
V.~Sze, Y.~Chen, T.~Yang, and J.~S. Emer, ``Efficient processing of deep neural
  networks: A tutorial and survey,'' \emph{Proceedings of the IEEE}, vol. 105,
  no.~12, pp. 2295--2329, Dec 2017.

\bibitem{Davis1980}
S.~B. Davis and P.~Mermelstein, ``Comparison of parametric representations for
  monosyllabic word recognition in continuously spoken sentences,''
  \emph{Acoustics, Speech and Signal Processing, IEEE Transactions on}, pp.
  357--366, 1980.

\bibitem{Cooley1965}
J.~Cooley and J.~Tukey, ``An algorithm for the machine calculation of complex
  fourier series,'' \emph{Mathematics of Computation}, vol.~19, no.~90, pp.
  297--301, 1965.

\bibitem{Rosenblatt1958}
F.~Rosenblatt, ``The perceptron: A probabilistic model for information storage
  and organization in the brain,'' \emph{Psychological Review}, pp. 65--386,
  1958.

\bibitem{Ioffe2015}
S.~Ioffe and C.~Szegedy, ``Batch normalization: Accelerating deep network
  training by reducing internal covariate shift,'' 2015, pp. 448--456.

\bibitem{CortexM4}
\emph{Cortex-M4: Technical Reference Manual}, ARM Limited, March 2010.

\bibitem{Haine17}
T.~Haine, Q.~Nguyen, F.~Stas, L.~Moreau, D.~Flandre, and D.~Bol, ``An 80-mhz
  0.4v ulv sram macro in 28nm fdsoi achieving 28-fj/bit access energy with a
  ulp bitcell and on-chip adaptive back bias generation,'' in \emph{ESSCIRC
  2017 - 43rd IEEE European Solid State Circuits Conference}, Sept 2017, pp.
  312--315.

\bibitem{CypressSRAM}
\emph{CY62126EV30 MoBL: 1-Mbit (64K x 16) Static RAM}, Cypress Semiconductor
  Corporation, 2017, rev. *P.

\bibitem{rosas2015impact}
F.~Rosas and C.~Oberli, ``Impact of the channel state information on the
  energy-efficiency of mimo communications,'' \emph{IEEE Transactions on
  Wireless Communications}, vol.~14, no.~8, pp. 4156--4169, 2015.

\bibitem{rosas2012modulation}
------, ``Modulation and snr optimization for achieving energy-efficient
  communications over short-range fading channels,'' \emph{IEEE Transactions on
  Wireless Communications}, vol.~11, no.~12, pp. 4286--4295, 2012.

\bibitem{rosas2016optimizing}
F.~Rosas, R.~D. Souza, M.~E. Pellenz, C.~Oberli, G.~Brante, M.~Verhelst, and
  S.~Pollin, ``Optimizing the code rate of energy-constrained wireless
  communications with harq,'' \emph{IEEE Transactions on Wireless
  Communications}, vol.~15, no.~1, pp. 191--205, 2016.

\bibitem{cui2005}
S.~Cui, A.~J. Goldsmith, and A.~Bahai, ``{Energy-constrained modulation
  optimization},'' \emph{IEEE Transactions on Wireless Communications}, vol.~4,
  no.~5, pp. 2349--2360, Sept. 2005.

\bibitem{cui2004}
------, ``{Energy-efficiency of MIMO and cooperative MIMO techniques in sensor
  networks},'' \emph{{IEEE} Journal on Selected Areas in Communications},
  vol.~22, no.~6, pp. 1089--1098, Aug. 2004.

\bibitem{Gustavsson2000}
J.~J.~W. M.~Gustavsson and N.~N. Tan, \emph{CMOS Data Converters for
  Communications}.\hskip 1em plus 0.5em minus 0.4em\relax Boston, MA: Kluwer,
  2000.

\bibitem{rosas2014optimizing}
F.~Rosas, G.~Brante, R.~D. Souza, and C.~Oberli, ``Optimizing the code rate for
  achieving energy-efficient wireless communications,'' in \emph{Wireless
  Communications and Networking Conference (WCNC), 2014 IEEE}.\hskip 1em plus
  0.5em minus 0.4em\relax IEEE, 2014, pp. 775--780.

\bibitem{rosas2013nakagami}
F.~Rosas and C.~Oberli, ``Nakagami-m approximations for multiple-input
  multiple-output singular value decomposition transmissions,'' \emph{IET
  Communications}, vol.~7, no.~6, pp. 554--561, 2013.

\bibitem{Siekkinen2012}
M.~Siekkinen, M.~Hiienkari, J.~Nurminen, and J.~Nieminen, ``How low energy is
  bluetooth low energy? comparative measurements with zigbee/802.15.4,'' in
  \emph{Wireless Communications and Networking Conference Workshops (WCNCW),
  2012 IEEE}, 2012, pp. 232--237.

\bibitem{Balankutty2010}
A.~Balankutty, S.-A. Yu, Y.~Feng, and P.~Kinget, ``{A 0.6-V Zero-IF/Low-IF
  Receiver With Integrated Fractional-N Synthesizer for 2.4-GHz ISM-Band
  Applications},'' \emph{IEEE Journal of Solid-State Circuits}, vol.~45, no.~3,
  pp. 538--553, March 2010.

\bibitem{IEEE-802.15.4short}
\emph{{Specifications for Local and Metropolitan Area Networks- Specific
  Requirements Part 15.4}}, IEEE Std. 802.15.4, 2006.

\end{thebibliography}
